\documentclass[conference]{IEEEtran}
\ifCLASSINFOpdf
\else
\fi
\usepackage{algorithm}

\usepackage{amssymb}
\usepackage{algorithm,algpseudocode}

\usepackage{subfig}

\usepackage{mathtools}
\usepackage{graphicx}
\usepackage{array}
\usepackage{float}
\usepackage{lipsum}
\usepackage{authblk}
\usepackage{authblk}

\begin{document}
%
\title{NeuRoute: Predictive Dynamic Routing for Software-Defined Networks}
%
%
%

\author{Abdelhadi Azzouni\thanks{abdelhadi.azzouni@lip6.fr}}

 \author[2]{Raouf Boutaba\thanks{E.E@university.edu}}
\author{Guy Pujolle\thanks{guy.pujolle@lip6.fr}}

\affil{LIP6 / UPMC; Paris, France  \{abdelhadi.azzouni,guy.pujolle\}@lip6.fr}

 \affil[2]{University of Waterloo; Waterloo, ON, Canada  rboutaba@uwaterloo.ca }
\maketitle

\begin{abstract}

This paper introduces NeuRoute, a dynamic routing framework for Software Defined Networks (SDN) entirely based on 
machine learning, specifically, Neural Networks. Current SDN/OpenFlow controllers use a default routing based on Dijkstra's algorithm
for shortest paths, and provide APIs to develop custom routing applications. NeuRoute is a controller-agnostic 
dynamic routing framework that (i) predicts traffic matrix in real time, 
(ii) uses a neural network to learn traffic characteristics and 
(iii) generates forwarding rules accordingly to optimize the network throughput. NeuRoute achieves 
the same results as the most efficient dynamic routing heuristic but in much less execution time.

%


\end{abstract}


 {\bf { \it keywords - }}
Routing, Machine Learning, Neural Networks, Software Defined Networking, Self Organizing Networks.

%

\IEEEpeerreviewmaketitle

\section{Introduction}

The modern Internet is experiencing an explosion of the Machine-to-Machine (M2M)
communications and Internet-of-Things (IoT) applications, 
in addition to other bandwidth intensive applications such as voice over
IP (VoIP), video conferencing and video streaming services. 
Thus leading to a high pressure on carrier operators to 
increase their network capacity in order to support all these applications
with an acceptable Quality of Service (QoS). The common practice to ensure a good QoS so far is 
to over-provision network resources. 
Operators over-provision a network so that capacity is based on peak traffic load estimates. 
Although this approach is simple for networks with predictable peak loads, 
it is not economically justified in the long-term. 


In addition, most ISP networks today use Shortest Path First (SPF) routing algorithms, 
namely the Open Shortest Path First (OSPF) \cite{ospf}. 
OSPF routes packets statically by assigning weights to links hence the routing
tables are recalculated only when a topology change occurs.
OSPF is a best effort routing protocol, meaning that when a packet experiences congestion, 
the routing subsystem cannot send it through an alternate path,
thus failing to provide desired QoS during congestion even when the total traffic
load is not particularly high.

Although OSPF has a QoS extension \cite{ospfqos}
that dynamically changes link weights based on measured traffic, 
it is still not implemented in the
Internet for two major reasons. First, changing the cost of a link
in one part of the network may cause a lot of routing updates
and in turn negatively affect traffic in a completely different
part of the network. This can be disruptive to many
(or all) traffic flows. Another problem concerns routing
loops that may occur before the routing protocol converges.
Therefore, in networks with distributed control plane, changing the
link cost is considered just as disruptive as link-failures.
On the other hand, without the possibility to differentiate between
traffic flows more granularly (not only based on destination IP
address), dynamic routing cannot positively contribute to
load balancing \cite{newapproach}.

The dynamic routing problem, also known as QoS routing 
or concurrent flow routing, is a case of Multi-commodity flow problem where flows 
are packets or traffic flows and the goal is to maximize the total network flow 
while respecting routing constraints such as load balancing the 
total network traffic or minimizing the traffic delay. 
Due to their high computational complexity, multi-commodity flow algorithms are rarely implemented 
in practice. 

There are many variants of the dynamic routing problem including the 
maximum throughput dynamic routing, the maximum throughput minimum cost 
dynamic routing and the maximum throughput minimum cost multicast dynamic routing.
In this work, we focus on the maximum throughput minimum cost unicast dynamic routing where given
a traffic demand matrix, the objective is to maximize the total throughput of the network 
while minimizing the cost of routing the total traffic knowing that each flow can be routed through 
only one end-to-end path. We present NeuRoute, a Neural Network based hyperheuristic 
that is capable of computing dynamic paths in real time. 
NeuRoute learns from a dynamic routing algorithm then 
imitates it achieving the same results but in only 25\% of its execution time. 
The basic motivation behind NeuRoute is that dynamic routing using traditional algorithmic solutions is 
not practical due to their high computational complexity. 
That is, at every execution round the routing algorithm uses measured link loads 
as input and performs a graph search to find the near optimal paths. 

The main contributions of this paper are summarized as follows: (i) We introduce for the first time an integral
routing system based on machine learning and detail its architecture, (ii) we detail the design of the
neural network responsible for matching traffic demands to routing paths and (iii) we evaluate our proposal against
an efficient dynamic routing heuristic and show our solution's superiority. 


The remainder of this paper is organized as follows: Section \ref{problemstatement} formally 
states the dynamic routing problem and discusses its most prominent heuristic solutions. Section \ref{neuroute} 
details NeuRoute design. In section \ref{eval}, we evaluate NeuRoute on real world network data and topology 
and we conclude the paper in section \ref{conclusion}

\section{The Dynamic Routing Problem} \label{problemstatement}

In this section, we formulate the maximum throughput minimum cost dynamic routing problem (MT-MC-DRP)
as a linear program, and then prove its NP-completeness.
The problem is equivalent to the known Unsplittable Constrained Multicommodity
Max-Flow-Min-Cost problem. We want to find routings for multiple
unicast flows which maximizes the aggregate flow in
a graph, while minimizing the routing-cost. By focusing on unsplittable 
multicommodity flow we exclude multipath routing where a flow can be split and routed through 
multiple end-to-end paths. 

We consider a software-defined network $G(V, L)$,
where $V$ is the set of SDN-enabled switch nodes, and $L$ is
the set of links that connect the switches where each link 
$l_{i,j}$ has a capacity $C(l)$. 
Each unicast flow $f$ has source and destination nodes
denoted $s_f$ and $d_f$ respectively, a requested traffic rate $R^f$ and a minimum necessary traffic rate $N^f$.
Let $r^f_{in}(v)$ and $r^f_{out}(v)$ denote the aggregate flow rate into/out
of node $v$ due to flow $f$, respectively. The traffic rate related to flow $f$ and flowing 
through link $l$ is denoted by $r^f(l)$.
Each ink has a routing 
cost denoted by $\Theta(l)$ that can represent any linear function of the traffic flowing on it
, i.e., delay, jitter, congestion probability or reliability.
We define an Admissible Routing as an assignment of flows to
the links in $G$, such that no capacity constraints are violated,
and flow-conservation applies at every node.
%
The MT-MC-DRP problem can be
stated as follows: Does there exist an admissible 
routing for the flows, where each flow receives its
requested rate $r^f$ while the total routing cost is minimized? 

%

\subsection{MT-MC-DRP As Two Linear Problems}

We formulate MT-MC-DRP as a succession of two linear problems (LPs): 
A Constrained-Maximum-Flow LP (CMaxF-LP) and a Constrained-Minimum-Cost LP (CMinC-LP).

\subsubsection{CMaxF-LP}
\begin{equation} \label{CMaxF-LP}
maximize (\sum_{f\in F}  r_{in}^f(d_f)) 
\end{equation}
subject to:
\begin{equation}
 r^f(l)  \geq 0  \quad \quad \forall f \in F, \forall l\in L^f 
\end{equation}
\begin{equation}
 r^f(l) \leq C(l) \quad \forall f \in F, \forall l \in L^f
\end{equation}
\begin{equation}
 \sum_{f\in F } r^f(l)\leq C(l) \quad \quad \quad \forall l \in L
\end{equation}
\begin{equation}
 r_{in}^f(v) = r_{out}^f(v) \quad \quad \quad \forall f \in F, \forall v \in V^f-\{s_f,d_f\}
\end{equation}
\begin{equation}
 r_{in}^f(s_f) = 0  \quad \quad \quad \forall f \in F
\end{equation}
\begin{equation}
 r_{out}^f(d_f) = 0 \quad \quad \quad \forall f \in F
\end{equation}
\begin{equation}
 r_{out}^f(s_f) \leq R^f  \quad \quad \quad \forall f \in F
\end{equation}
\begin{equation}
 r_{out}^f(s_f) \geq N^f  \quad \quad \quad \forall f \in F
\end{equation}

\subsubsection{CMinC-LP}
\begin{equation} \label{CMinC-LP}
minimize (\sum_{f\in F} \sum_{l\in L} r^f(l) \times \Theta(l)) 
\end{equation}
subject to:
\begin{equation}
 r_{out}^f(s_f) = \Pi^f +/- \epsilon \quad \quad \forall f\in F, \forall  l\in L^f
\end{equation}
\begin{equation}
 \sum_{f\in F } r^c(l)\leq  C(l) \quad \quad \quad\forall l \in L
\end{equation}
\begin{equation}
 r_{in}^f(v) = r_{out}^f(v) \quad \quad \quad \forall f \in F, \forall v \in V'
\end{equation}
\begin{equation}
 r_{in}^f(s_f) = 0  \quad \quad \quad \forall f \in F
\end{equation}
\begin{equation}
 r_{out}^f(d_f) = 0 \quad \quad \quad \forall f \in F
\end{equation}

\textit{Theorem.} The Maximum Throughput Minimum Cost Dynamic Routing Problem as 
presented above is NP-hard. \\

\textit{Proof.} refer to \cite{overtime} \cite{maxminrouting} $\square$

\subsection{Heuristic Solution for The MT-MC-DRP} \label{BH}
Due to its NP-completeness, an exact solution for the MT-MC-DRP 
as defined above is not practical to be implemented in the network controller. 
It is more practical to design an approximate but fast 
solution. Therefore, a major research effort was put into designing efficient 
fully polynomial-time approximation schemes (FPTAS) for multicommodity flow problems 
including max flow min cost multicommodity problem.
A fully polynomial-time approximation scheme for a flow maximization problem 
is an algorithm that,
given an accuracy parameter $\epsilon > 0$, computes, in polynomial time
in the size of the input and $1/\epsilon$, a solution with an objective value 
within a factor of $(1 - \epsilon)$ of the
optimal one \cite{fasf}. The multicommodity problem literature
has a rich body of work providing FPTASes. In this work, we use the novel method 
proposed in \cite{fasf} as a baseline heuristic to solve the MT-MC-DRP. We also refer to the same 
paper for more literature on other existing heuristics.

\section{System Design} \label{neuroute}

\begin{figure}[h] 
\centering
   \includegraphics[scale=0.3]{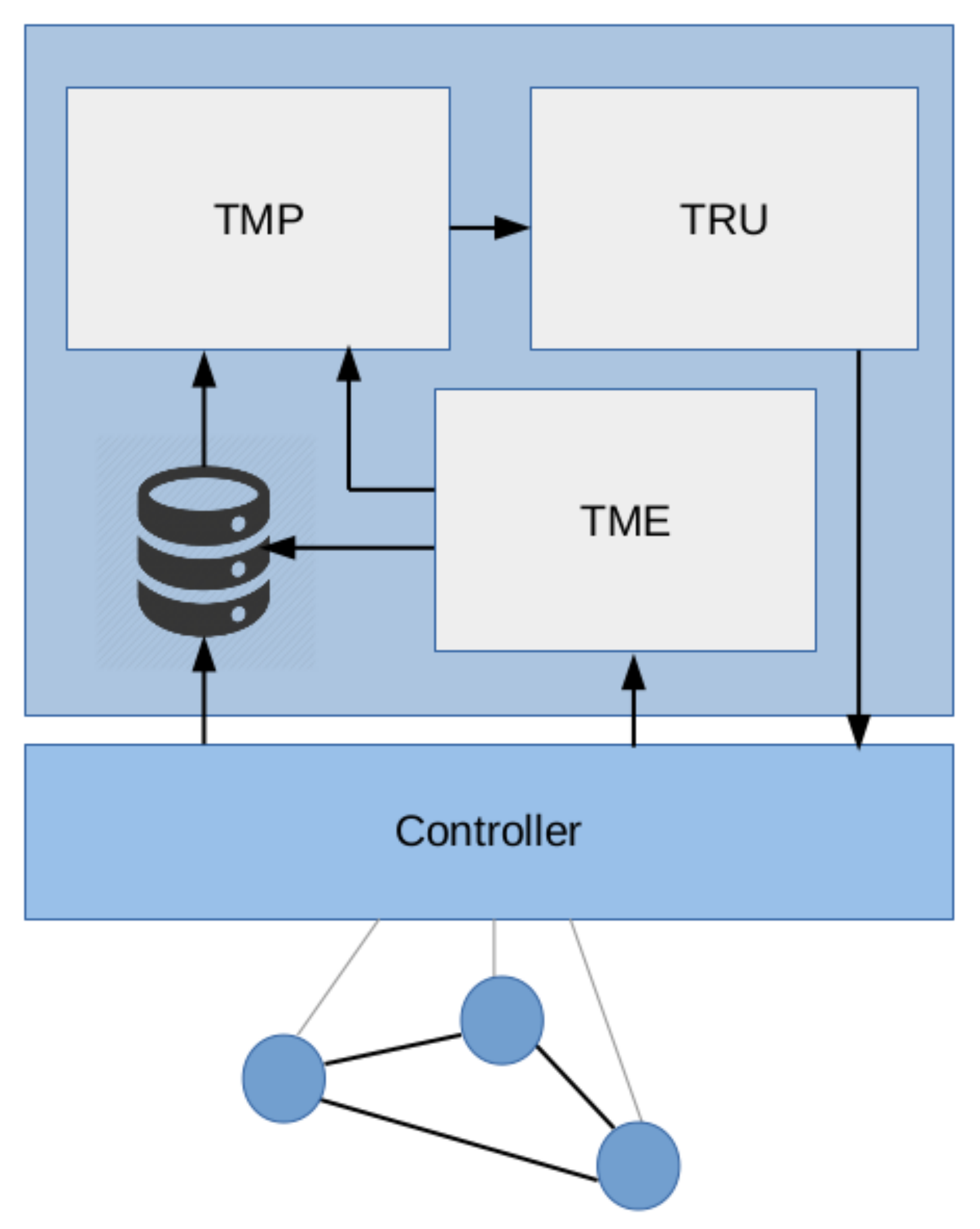}
   \caption{\label{fig:lldppacket} NeuRoute architecture}\label{nrdesign}
\end{figure}

As shown in figure \ref{nrdesign}, NeuRoute is designed as an integral 
routing application for the SDN controller. 
NeuRoute is composed of three key components: a Traffic Matrix Estimator (TME), 
a Traffic Matrix Predictor (TMP) and a Traffic Routing Unit (TRU). In this paper focus on and 
detail the TRU but also describe briefly the two other components for the sake of completeness.

\subsection{Traffic Matrix Estimator}

As mentioned earlier,detailed design of
the traffic matrix (TM) estimator is out of the scope of this paper. 
Here we only motivate the need for 
a traffic matrix estimator and define its interfaces 
with the rest of NeuRoute components.

A network TM presents the traffic volume between all pairs
of origin-destination (OD) nodes of the network at a
certain time t. The nodes in a traffic matrix can be Points-
of-Presence (PoPs), switches, routers or links.
In OpenFlow SDNs, the controller leverages packet\_in messages 
to build a global view of the network.
When a new flow arrives to a switch, it is matched against forwarding rules to determine 
a forwarding path for it. If the flow does not match any rule, the switch forwards the first packet or only 
the packet header to the controller. In addition, the controller can query switches for 
packet counts that track the number of packets and bytes handled by the switch. 
However, the number of packet\_in and the number of 
controller queries, necessary for a near real-time measurement, 
increases rapidly with a large number of switches and flows, making
this measurement mechanism not practical.
Also, there is a chance that by the time the controller receives the message, 
the values of the counters become out of date and do not reflect the 
near real-time state of the switch anymore. 
These and a number of other issues listed in \cite{opensample} call for an 
efficient measurement mechanism to capture traffic matrix in near real-time. In its current implementation, NeuRoute 
uses a variant of a recent proposal called openMeasure \cite{openmeasure} to estimate traffic matrix.

\subsection{Traffic Matrix Predictor}

\begin{figure}[h] 
\centering
   \includegraphics[scale=0.4]{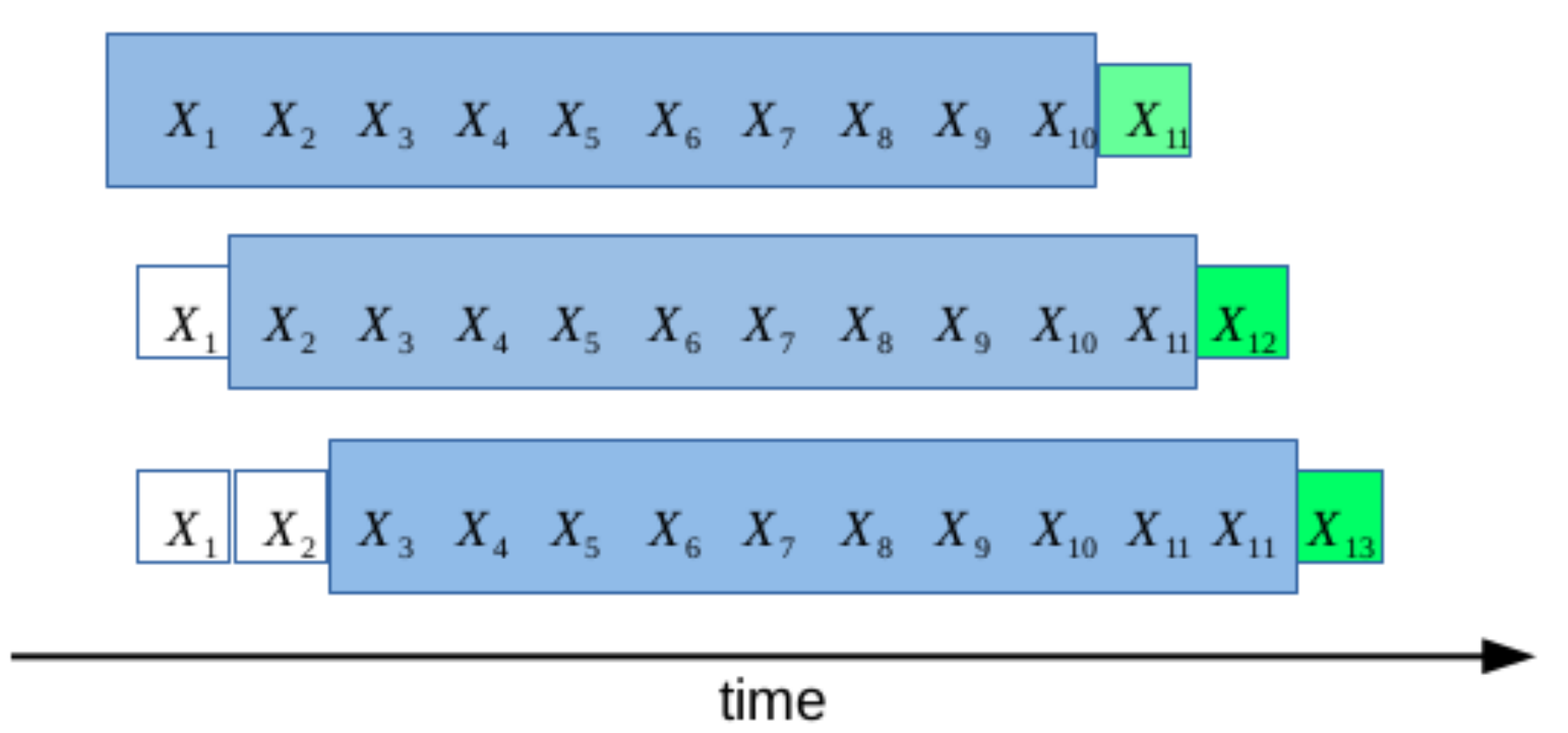}
   \caption{\label{fig:lldppacket} Traffic Matrix Prediction Over Time}\label{window}
    \vspace{-1em}
\end{figure}

Network Traffic Matrix prediction refers to the problem of estimating future network traffic from the
past and current network traffic data. 
Internet traffic is known to be self-similar enabling it to be 
predictable with high accuracy \cite{selfsimilar}. 
NeuRoute's Traffic Matrix Predictor (TMP) uses a 
Long Short Term Memory Recurrent Neural Network (LSTM-RNN)
described in \cite{lstmtm}. Figure \ref{window} shows 
the sliding prediction window where at each time instant $t$, the TMP takes 
a fixed size set of achieved traffic matrices as input and outputs the traffic matrix of time instant $t+1$

Prediction using NNs involves two phases: a) the training
phase and b) the test (prediction) phase. During the training
phase, the NN is supervised to learn from the data by 
presenting the training data at the input layer and dynamically
adjusting the parameters of the NN to achieve the desired
output value for the input set. The most commonly used
learning algorithm to train NNs is called the backpropagation
algorithm. The underlying idea is to propagate 
the error backward, from the output to the input, where the
weights are changed continuously until the output error falls
below a preset value. In this way, the NN learns correlated
patterns between input sets and the corresponding target val-
ues. The prediction phase represents the testing of the NN. A
new unseen input is presented to the NN and the output is
calculated, thereby predicting the outcome of new input data.

\subsection{Traffic Routing Unit}

The core component of the NeuRoute system is the Traffic Routing Unit (TRU) which is responsible of selecting optimal 
routes based on the  
predicted traffic matrix. 
TRU is based on the supervised learning approach where an agent is trained
to infer a function from labeled training data. 
It consists of a Deep Feed Forward Neural Network that learns 
to match traffic demands to routing paths by observing the output of a 
heuristic, that we call the Baseline Heuristic (BH). In this paper we present our 
experimentations with a BH that is built following the algorithm discussed in section \ref{BH}.


To bootstrap, only the TME is activated to continuously 
provide the BH with timely estimated traffic matrices. Copies of these estimated traffic matrices are stored 
to be used later on by the TMP and the TRU.
NeuRoute collects the output of the BH for a period of time that can be configured 
based on the desired performance. Once enough BH-output data is gathered, NeuRoute's components, TMP and TRU,
are fired up. The TMP uses the stored history of estimated traffic matrices to predict 
the future traffic matrix, continuously as detailed in \cite{lstmtm}. On the other hand, 
the TRU takes the BH output data and the stored history of estimated 
traffic matrices along with corresponding Network States (NSs) as input to train its routing neural network. 
Each tuple (NS+traffic matrix, BH output)
constitutes one learning sample for the TRU. $NS$ at a time instant $t$ (or $NS_t$) is the set of all 
links available capacities and links costs at time instant $t$ (links costs usually do not change frequently). 
Once the learning phase is done 
(within a few seconds to a few minutes depending on the volume of data and desired performance), 
the trained model is fired up to route new traffic flows. 
The reason why we predict the traffic matrix
is that the real-time measurement of traffic matrix is not practical and by
the time the controller gets the measured information, the flows to be routed are already
on their way on the existing paths, before even the controller computes the new paths.
In the following, we detail the design elements and the design challenges of TRU.

\subsubsection{Deep Feed Forward Neural Networks}

\begin{figure}[h] 
\centering
   \includegraphics[scale=0.35]{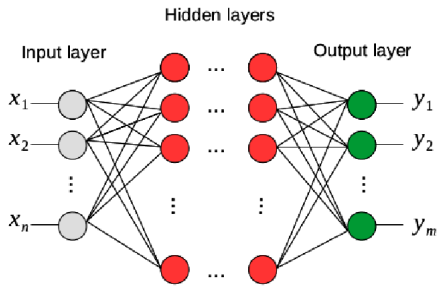}
   \caption{\label{fig:lldppacket} Deep Feed Forward Neural Network}\label{ffnnn}
\end{figure}

Deep neural networks are currently the most successful machine learning 
technique for solving a variety of tasks including language translation, 
image classification and image generation. 
TRU is similar to an image classifier that has a set of images in input 
and tries to find a function that matches these images to a set of classes. 
In the routing case, the traffic matrices are the images and the routing paths represent the output classes.
The deep neural network used in TRU is presented in figure \ref{ffnnn}.
It takes a traffic matrix and an NS instance
as input and matches them to a unique path $y_i$ as output. 

In a deep feed forward network, the information flows only forward through the network from the input nodes, 
through the hidden nodes to the output nodes, with no cycles or loops. 
Each node has an activation function which acts like a threshold for the node to fire up:
A node $n$ produces a value for its output nodes only if the weighted sum of the 
input values of $n$ is equal or exceeds the threshold.
Each edge has a weight and permits transfer of value from node to node.

\textbf{Learning Algorithm. } We use the Backpropagation learning algorithm that was first introduced in the 70s and now
is the most widely used algorithm for supervised learning in
deep feed-forward networks. 
The goal is to make the network learn some target function, in our case, matching 
traffic matrices to routing paths.
The basic idea of the algorithm 
is to look for the minimum of the error function
in weight space by repeatedly 
applying the chain rule to compute the influence of each weight in
the network with respect to the error function: 
The output values of the network are compared with the learning sample (correct answer) 
to compute the value of the error function. 
The calculated error is then fed back through the network and used  
to adjust the weights of each connection in order to reduce the value of the error function by some small amount.
After repeating this process for a sufficiently large number of training cycles, 
the network will usually converge to some state where the error is small enough. 
In other words, we say that the network 
has learned the target function to some extend. We refer to \cite{backprop}
for more details about the algorithm.
%
%

%
\textbf{Optimization Algorithm. }
In this work, we use Adam (short for Adaptive Moment Estimation) optimizer, 
one of the most adopted optimization algorithms among 
deep learning practitioners for applications in computer vision and natural language processing.
Adam optimizer is an improvement of the gradient descent algorithm 
that can yield quicker convergence in training deep networks \cite{adam}.

\textbf{Learning Rate. } The learning rate determines how quickly 
or how slowly we want the network weights to be updated (by the backpropagation algorithm). 
In other words, how quickly 
or how slowly we want the network to forget learned features and learn new ones. 
Picking a learning rate is 
problem dependent since the optimum learning rate can differ based 
on a number of parameters including epoch size, number of learning iterations, 
number of hidden layers and/or neurons and number and format of the inputs. 
Trial and error is often used in order to determine the ideal learning condition for each problem studied.
We describe our empirical approach for choosing the learning rate in the implementation section \ref{eval}.

\subsubsection{Input Pre-Processing and Normalization}

The input (NS+traffic matrix) are merged into one single vector of 
numbers then normalized by dividing all numbers by the greatest number. The result is 
a vector of numbers ranging between 0 and 1. This normalization is a good practice 
that can make training faster and reduce the chance of getting stuck in local optima \cite{normalize}.

\subsubsection{Routing Over Time}
At each time instant $t$, the TRU's trained model takes predicted traffic matrix of time instant $t+1$ ($TM_{t+1}$) and 
corresponding NS as input. The model function is applied and the output is a set of path probabilities where the highest 
value indicates the best routing path. TRU then sends the chosen path to the controller in order to be installed in 
switches as flow rules. By the time $t+1$, when the flows arrive, the forwarding rules are already installed which minimizes 
considerably the network delay.  \\

Matching traffic matrices and network states to routing paths is similar to classifying a stream of frames in a video, witch
is not a common and well studied problem since the usual image classification is applied
to individual images. Besides tweaking the neural network architecture and parameters to obtain a high classification performance, 
there are two unique challenges that arise in our problem:
\begin{itemize}
 \item The runtime performance of the trained model is critical and needs to be optimized 
 to perform continuous routing over time. We achieve high performance by keeping the predicted traffic matrices in memory 
 before feeding them to the LRU's neural network. 
\item Unlike images and videos, there is no 
camera bias in traffic matrices (Camera bias refers to the fact that in many images and videos, 
the object of interest often occupies the center region), hence it is not possible to work around 
resolutions to optimize training time as it was done in \cite{largescalevideo}.
\end{itemize}

\section{Implementation and Evaluation} \label{eval}

\begin{figure}[h] 
\centering
   \includegraphics[scale=0.2]{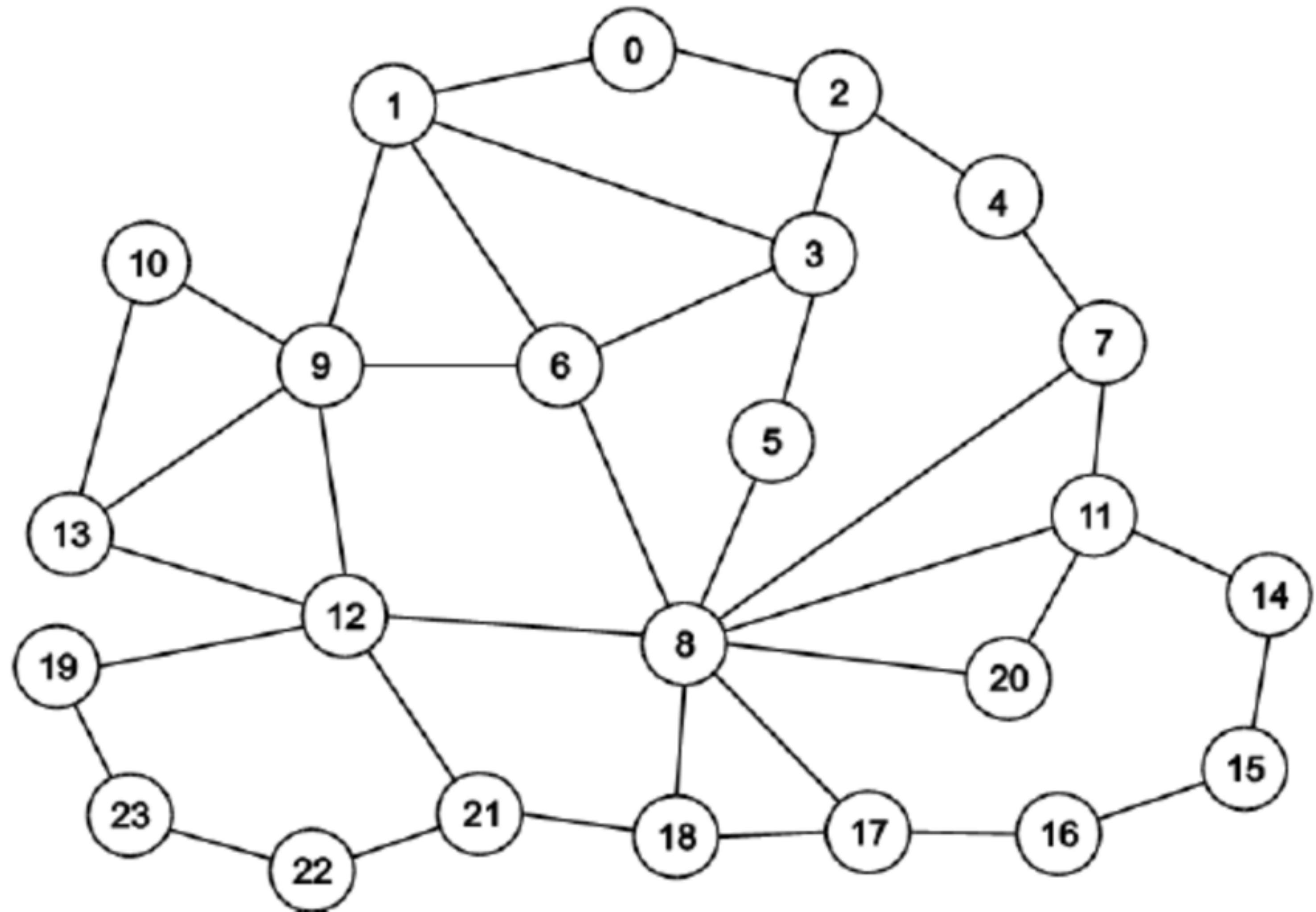}
   \caption{\label{fig:lldppacket} G\'EANT2 Network Topology \cite{geant}}\label{geant}
\end{figure}

We implemented NeuRoute as a routing application on top of POX controller \cite{pox}. 
The TRU's neural network is implemented using Keras library \cite{keras} on top of Google's TensorFlow 
machine learning framework \cite{tf}. We have chosen the G\'EANT network topology for our testbed as 
G\'EANT's traffic matrices are already available online \cite{geanttms}. 
We implemented the G\'EANT topology (shown in figure \ref{geant}) as an SDN network using Mininet \cite{mininet} setting 
link capacities at 10Mbps. We use link delay as the cost function with 2ms delay per link. 

\textbf{Data generation. }
In order to generate the learning data, we applied the BH on the testbed described above with G\'EANT's traffic matrices as input. 
We obtained a data set of 10000 samples (traffic matrix+network state, near optimal path) that we split to training data set of 7000 samples
and test data set of 3000 samples. 

\textbf{The neural network architecture. } Determining the neural network architecture is 
problem dependent, hence we adopted an empirical approach 
to determine the number of hidden layers and the size of each hidden layer. 
We measured the training time and the learning performance (G\'EANT traffic matrices + related network states as input and 
the results of the BH as output) for different numbers of hidden layers and different hidden layer sizes.
This allowed us to pick an optimal number of hidden layers of 6 with 100 nodes per hidden layer.
Note that we choose the architecture parameters based on the measured learning performance, and we stop experimenting when 
the training time becomes too long. 

%
%
%
%
%


\begin{figure}[h]
    \centering
    \subfloat[MSR over number of hidden layers \label{fig:msrbynhiddenlayers}] {{\includegraphics[scale=0.3]{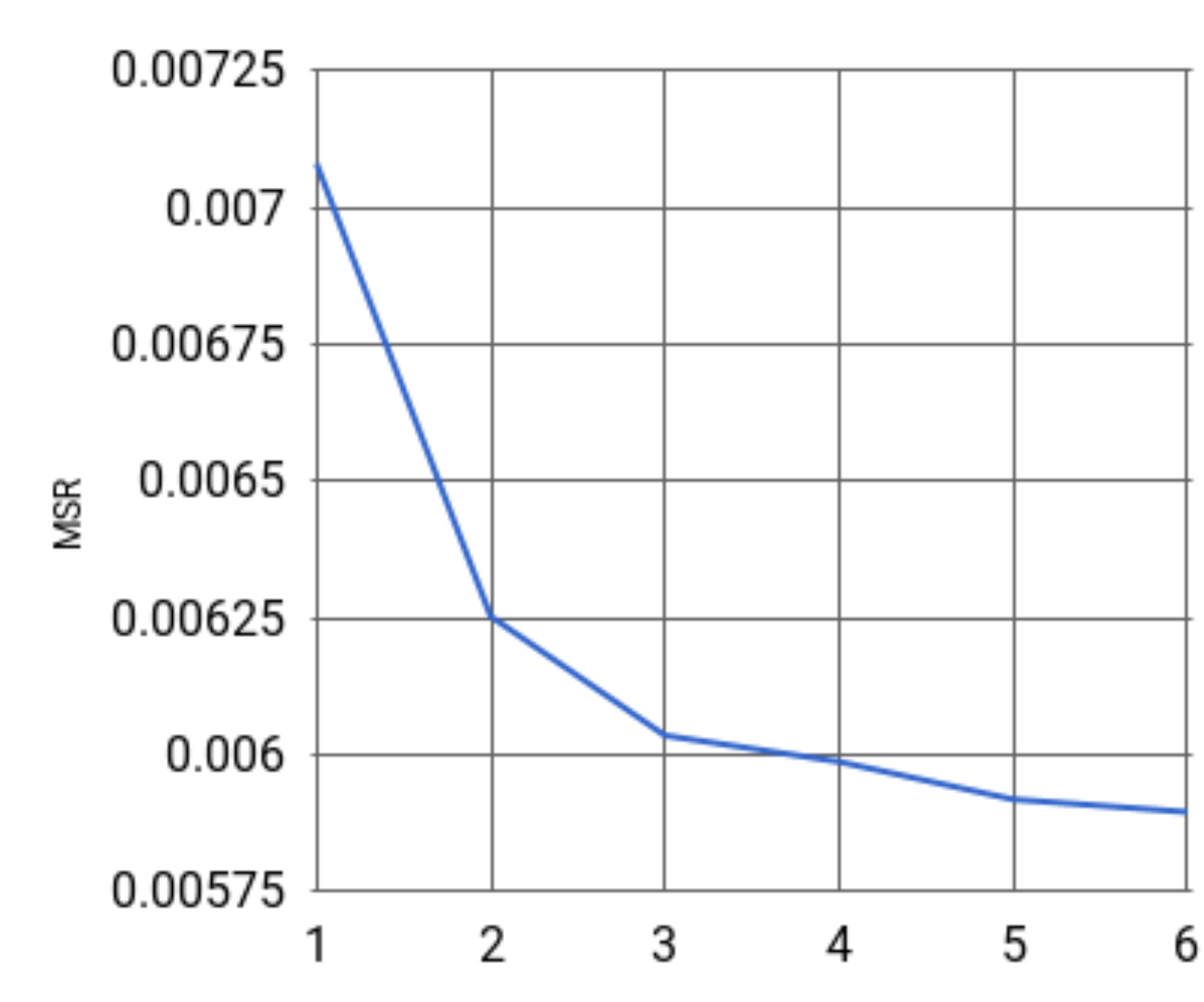} }} 
    \hfill
    \subfloat[Training time over number of hidden layers \label{fig:ltbynhiddenlayers}]{{\includegraphics[scale=0.32]{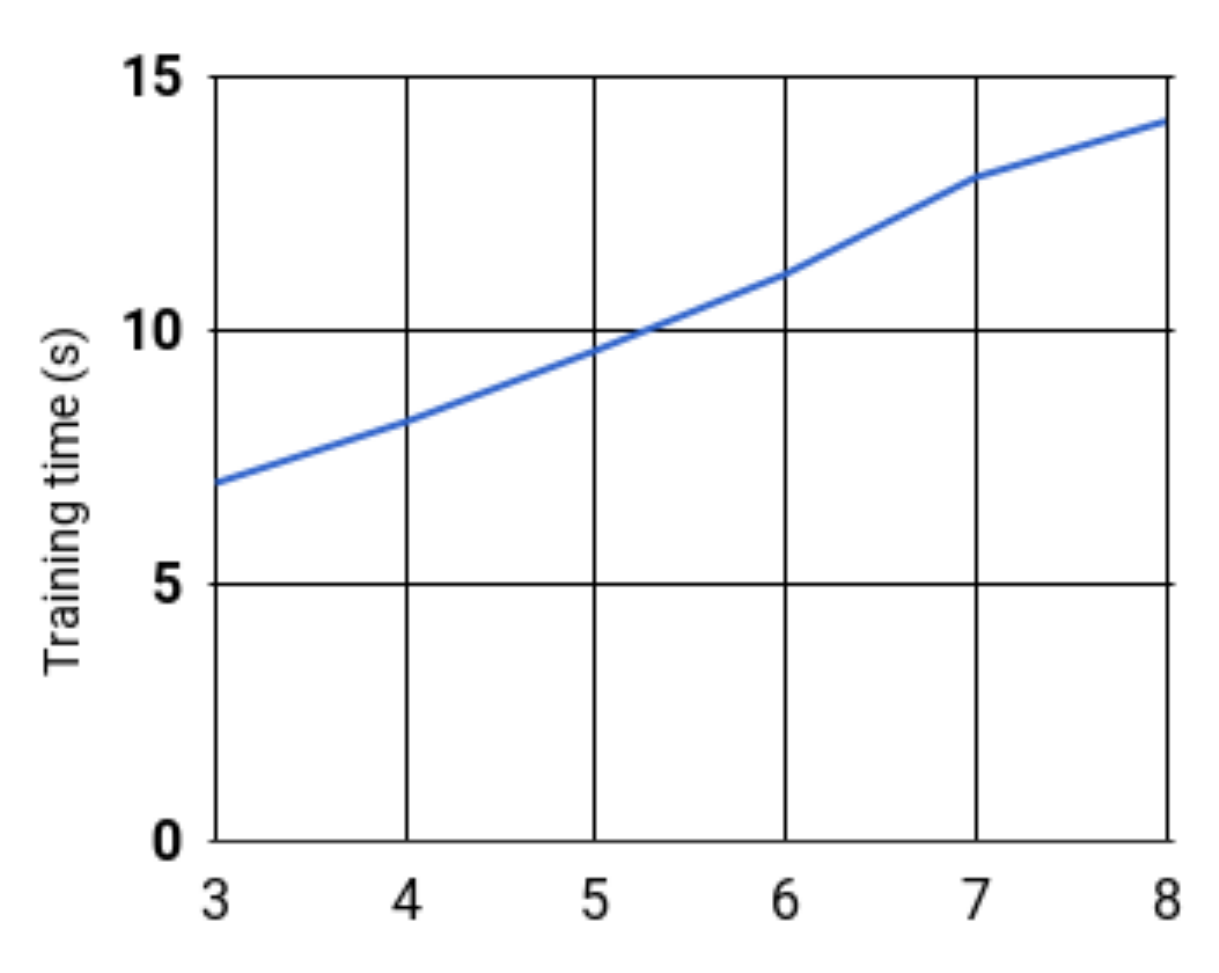} }} %
    \caption{Picking the number of hidden layers}%
    \label{fig:bynhiddenlayers}%
\end{figure}

\begin{figure}[h]
    \centering
    \subfloat[MSR over number of hidden nodes \label{fig:msrbynhiddennodes}]{{\includegraphics[scale=0.31]{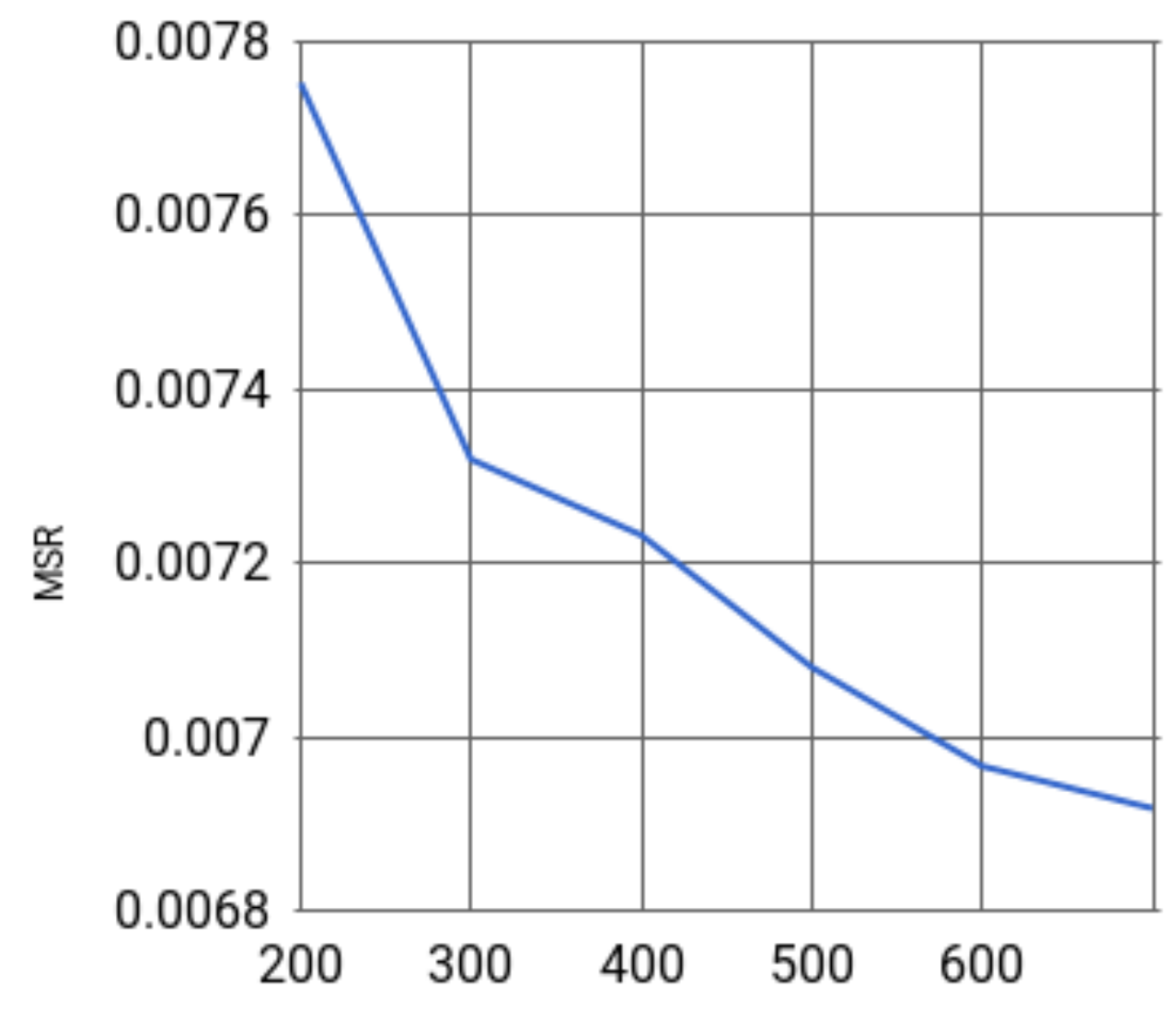} }}  %
    \hfill
    \subfloat[Training time over number of hidden nodes \label{fig:ltbynhiddennodes}]{{\includegraphics[scale=0.31]{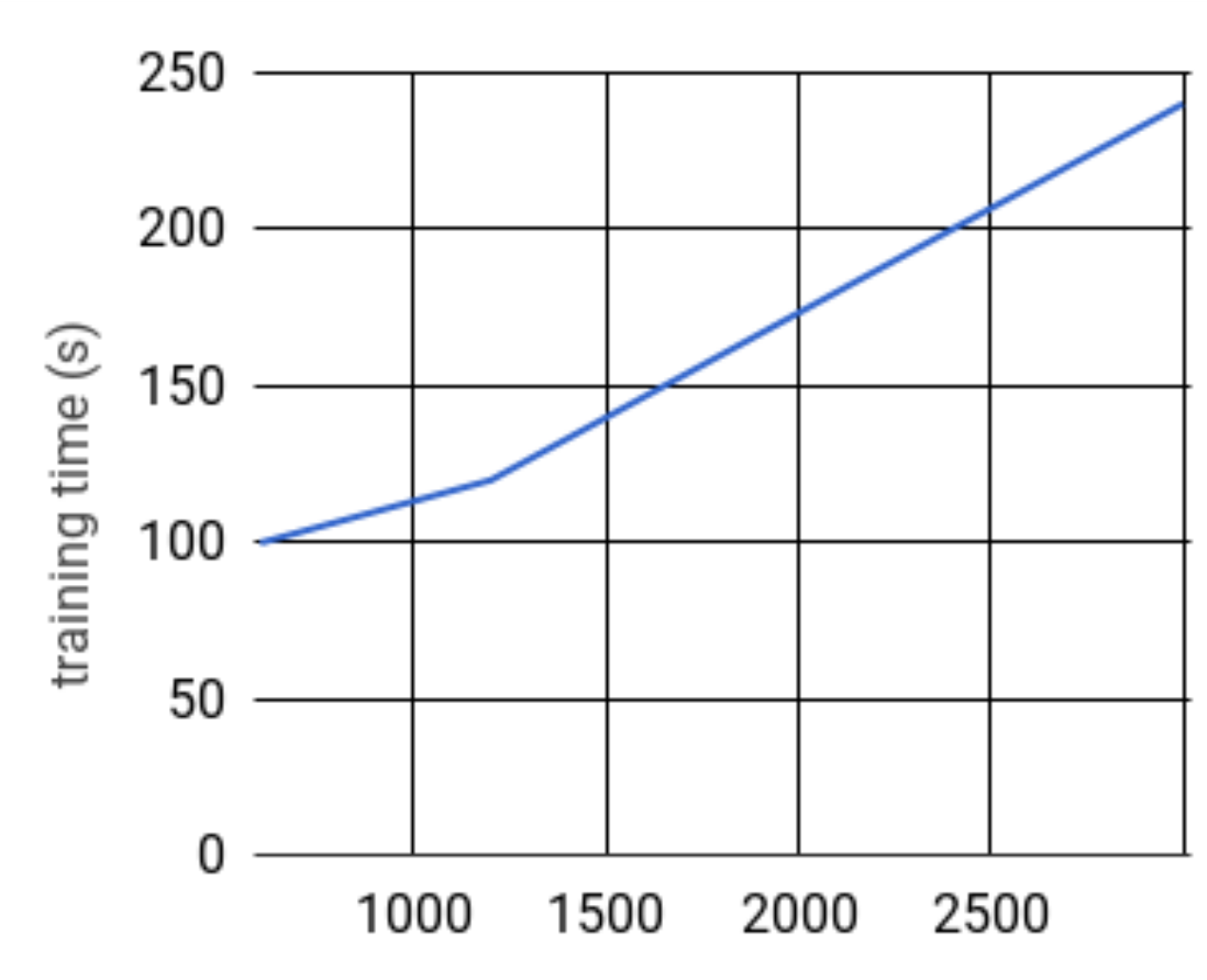} }}  %
    \caption{Picking the number of hidden nodes}%
    \label{fig:bynhiddennodes}%
\end{figure}

Figure \ref{fig:msrbynhiddenlayers} depicts the measured MSR over different numbers of hidden layers. 
The MSR diminishes at high numbers of hidden layers (deep network) but figure \ref{fig:ltbynhiddenlayers}
shows that the deeper is the network the longer it takes to train it. To select a good compromise, we fix 
the training time to 2 minutes. This training time corresponds to a depth of 6 hidden layers.

Similarly, figure \ref{fig:msrbynhiddennodes} shows that the MSR diminishes at higher network sizes but 
the training time goes up as figure \ref{fig:ltbynhiddennodes} shows. We fix again the training time to 2 minutes and obtain the corresponding 
hidden nodes number of 600, or 100 nodes per hidden layer. Note that a 2 minutes training time is not too long but is chosen 
proportionally to the size of the data set. Larger data sets may take hours or days to train.

\textbf{Data preparation. }
We prepared the input data as follows: we split the total learning data into batches of size 100 each. Each input sample
is a vector of size $506+38=544$, 506 being the size of a vector representing one traffic matrix of 23 nodes (23*22) and 38 being the 
number of links in the G\'EANT topology, which is equal to the size of one network state vector. The output vector is of size $23*22*5$ with 
23*22 being the number of origin-destination (OD) pairs and we arbitrarily fix the number of possible paths per OD pair to 6.


\textbf{The learning rate. } Like the neural network's architecture, the learning rate is problem dependent. 
Our approach is to start with a high value and go down to lower values, recording the learning performance and 
training time for every learning rate value. 


  \begin{figure}[h] 
\centering
   \includegraphics[scale=0.3]{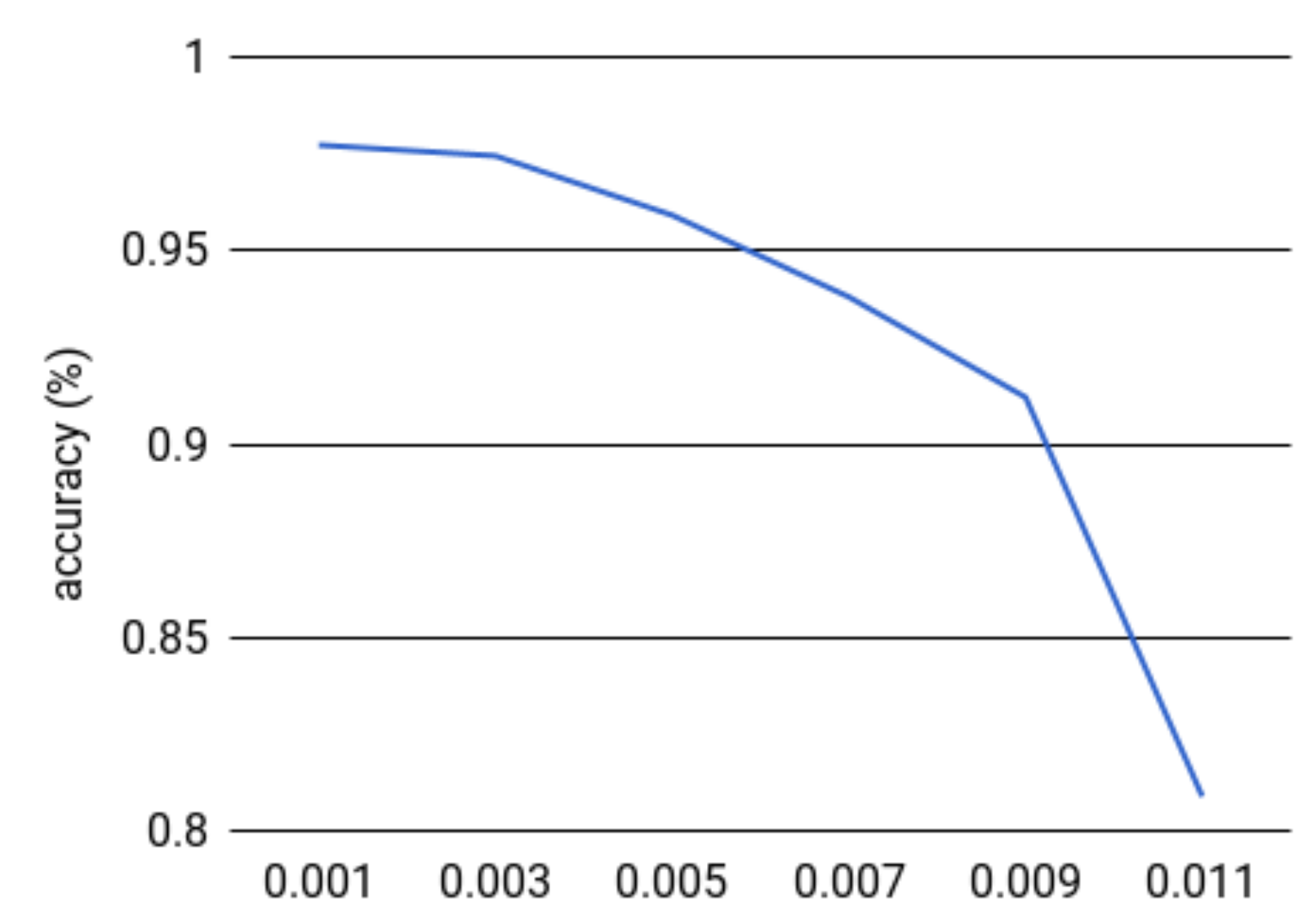}
   \caption{\label{fig:msrbylearningrate} MSR over different learning rate values }\label{./pdf/msrbynhiddenlayers.eps}
\end{figure}

Figure \ref{fig:msrbylearningrate} depicts the MSR variation over different learning rate values.
The training time does not change for different learning rates (5s per epoch).  

\textbf{The overfitting problem. }
Overfitting is a serious problem that occurs when training a neural network on limited data. 
It happens when a model learns the detail and noise in the training data to the extent that it negatively impacts 
its performance on new data. This means that the noise or random fluctuations in the training data is picked
up and learned as features by the model. The problem is that these features do not apply to new data and negatively impact 
the model's ability to generalize.
Various methods have been proposed to avoid or reduce overfitting, including stopping the training as soon as
performance on a validation set starts to get worse, introducing weight penalties of various
kinds such as L1 and L2 regularization and Dropout \cite{nitish}.
In this work, we use the Dropout technique which is proven to be the most effective \cite{nitish}.
Dropout is a technique that addresses both these issues. It prevents overfitting and provides 
a way of approximately combining exponentially many different neural network
architectures efficiently. The term “dropout” refers to dropping out units (hidden and 
visible) in a neural network.

\begin{figure}[h] 
\centering
   \includegraphics[scale=0.33]{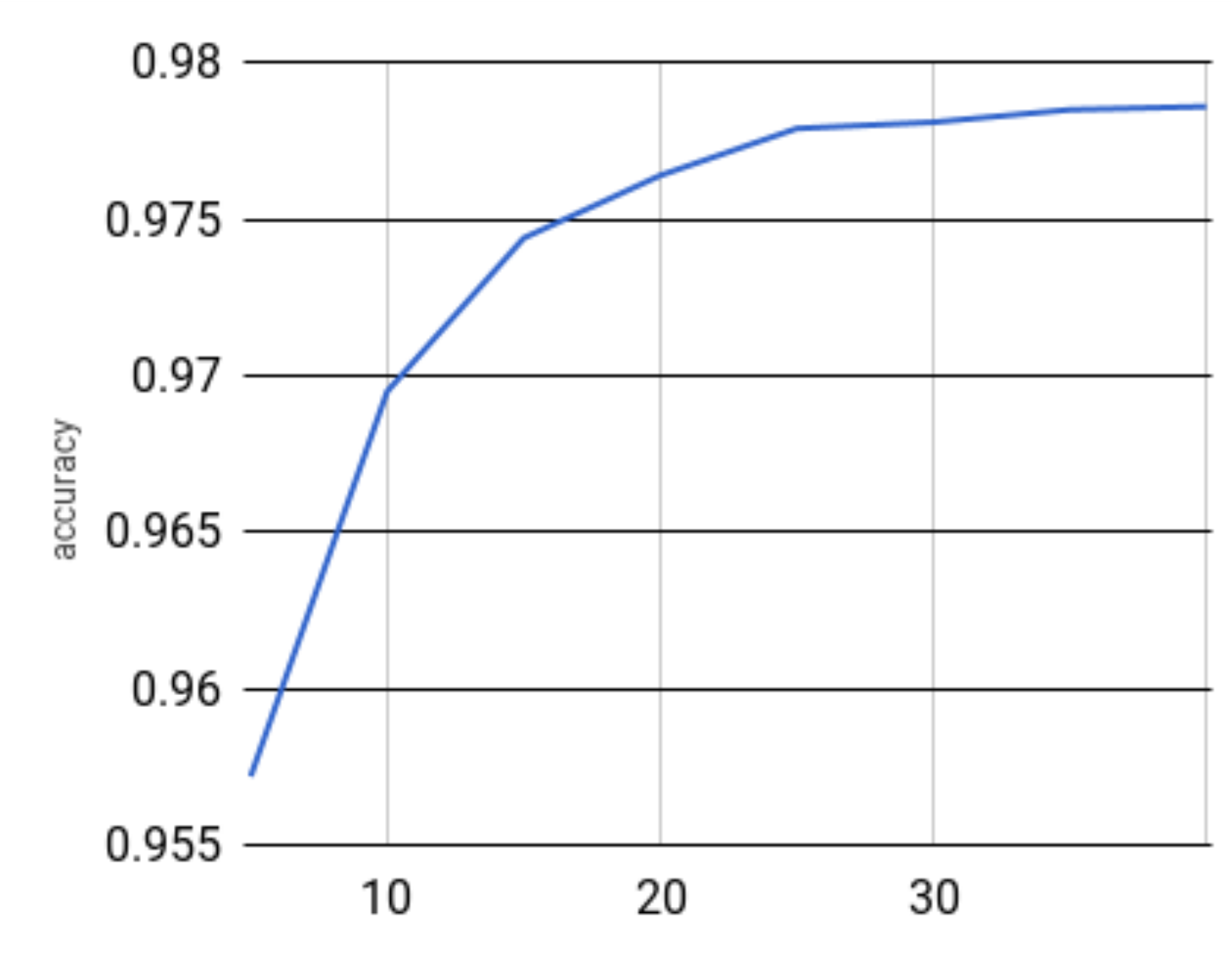}
   \caption{\label{fig:lldppacket} MSR over number of training epochs}\label{accuracy}
\end{figure}

\textbf{Evaluation of TRU. }
Finally, we applied the trained model on the test data and recorded the accuracy (number of correctly chosen paths from the test set)
over number of training 
epochs in figure \ref{accuracy}. 
One epoch is a one complete training pass over the whole training data set where each epoch takes roughly 2s to complete.
Figure \ref{accuracy} shows that the model picks the near optimal path learned from the BH with an estimated error of less than 0.05\% when trained well
(3min of training is enough to reach this error rate). Furthermore, the trained model executes and finds the near optimal path in 30ms compared to
the BH execution time of 120ms.

 
%
\section{Related Work} \label{relatedwork}

The authors of paper \cite{metalayer} propose a machine learning meta-layer composed of multiple modules. 
Each module works only for one OD pair. The proposed scheme is however not practical since the 
number of OD pairs (hence the number of neural networks associated) explodes in large networks.
Knowing that each neural network is trained separately and each trained model operates separately,
this approach does not capture the relations between ODs requests that arrive at the same time. 
It is also much more complicated to implement and computationally expensive than our approach.

\section{Conclusion} \label{conclusion}

In this paper, we introduced NeuRoute, a machine learning based dynamic routing framework for SDN. NeuRoute learns 
a routing algorithm and imitates it with higher performance. We implemented NeuRoute as a routing application on top of Pox 
Controller and performed proof of concept experiments that showed our solution's superiority compared to an efficient dynamic 
routing heuristic.
Experiments on larger data sets are being conducted and will be presented in 
a future work along with more details about the system.

\end{document}